\documentclass[aps,prl,twocolumn,superscriptaddress,showpacs]{revtex4-1}

\usepackage{graphicx,amsmath,amssymb,bm}
\usepackage[T1]{fontenc}

\newcommand{\bra}[1]{\left\langle #1\right|}
\newcommand{\ket}[1]{\left| #1\right\rangle}
\newcommand{\kev}{\,\mathrm{keV}}
\newcommand{\mev}{\,\mathrm{MeV}}

\newcommand{\gevi}{\,\mathrm{GeV}^{-1}}

\newcommand{\fmi}{\,\mathrm{fm}^{-1}}
\newcommand{\fmiq}{\,\mathrm{fm}^{-3}}

\begin{document}

\title{To which densities is spin-polarized neutron matter a weakly interacting Fermi gas?}

\author{T.\ Kr\"uger}
\email[E-mail:~]{thomas.krueger@physik.tu-darmstadt.de}
\affiliation{Institut f\"ur Kernphysik, Technische Universit\"at Darmstadt, 
64289 Darmstadt, Germany}
\affiliation{ExtreMe Matter Institute EMMI, GSI Helmholtzzentrum f\"ur 
Schwerionenforschung GmbH, 64291 Darmstadt, Germany}
\author{K.\ Hebeler}
\email[E-mail:~]{kai.hebeler@physik.tu-darmstadt.de}
\affiliation{Institut f\"ur Kernphysik, Technische Universit\"at Darmstadt, 
64289 Darmstadt, Germany}
\affiliation{ExtreMe Matter Institute EMMI, GSI Helmholtzzentrum f\"ur 
Schwerionenforschung GmbH, 64291 Darmstadt, Germany}
\author{A.\ Schwenk}
\email[E-mail:~]{schwenk@physik.tu-darmstadt.de}
\affiliation{Institut f\"ur Kernphysik, Technische Universit\"at Darmstadt,
64289 Darmstadt, Germany}
\affiliation{ExtreMe Matter Institute EMMI, GSI Helmholtzzentrum f\"ur
Schwerionenforschung GmbH, 64291 Darmstadt, Germany}

\begin{abstract}

We study the properties of spin-polarized neutron matter at
next-to-next-to-next-to-leading order in chiral effective field
theory, including two-, three-, and four-neutron interactions. The
energy of spin-polarized neutrons is remarkably close to a
non-interacting system at least up to saturation density, where
interaction effects provide less than 10\% corrections. This shows
that the physics of neutron matter is similar to a unitary gas well
beyond the scattering-length regime. Implications for energy-density
functionals and for a possible ferromagnetic transition in neutron
stars are discussed. Our predictions can be tested with lattice QCD,
and we present results for varying pion mass.

\end{abstract}


\maketitle

\textit{Introduction.}-- Due to the large neutron-neutron scattering
length, the physics of neutron matter exhibits properties similar to a
unitary Fermi gas~\cite{Carl03largea,Schw05dEFT,Carl12PTEP}. The
energy of neutron matter is approximately $0.4$ times the energy of a
free Fermi gas, and neutrons form an $S$-wave superfluid for densities
almost up to saturation density, for recent reviews see
Refs.~\cite{Carl12PTEP,Geze14pairing}. These benchmark results,
combined with the possibility to simulate low-density neutron matter
with ultracold atoms near a Feshbach resonance~\cite{Bloc08RMP}, have
lead to the inclusion of ab initio results for neutron matter into
modern energy-density functionals for nuclei~\cite{Bulg10LNP,Gori13nm}
and into predictions for neutron
stars~\cite{Hebe10PRL,Gand12nm}. Neutron matter is also interesting
theoretically, because all many-body forces among neutrons are
predicted in chiral effective field theory (EFT) to
next-to-next-to-next-to-leading order
(N$^3$LO)~\cite{Hebe10nmatt,Tews13N3LO}.

\begin{figure}[b]
\begin{center}
\includegraphics[width=0.9\columnwidth,clip=]{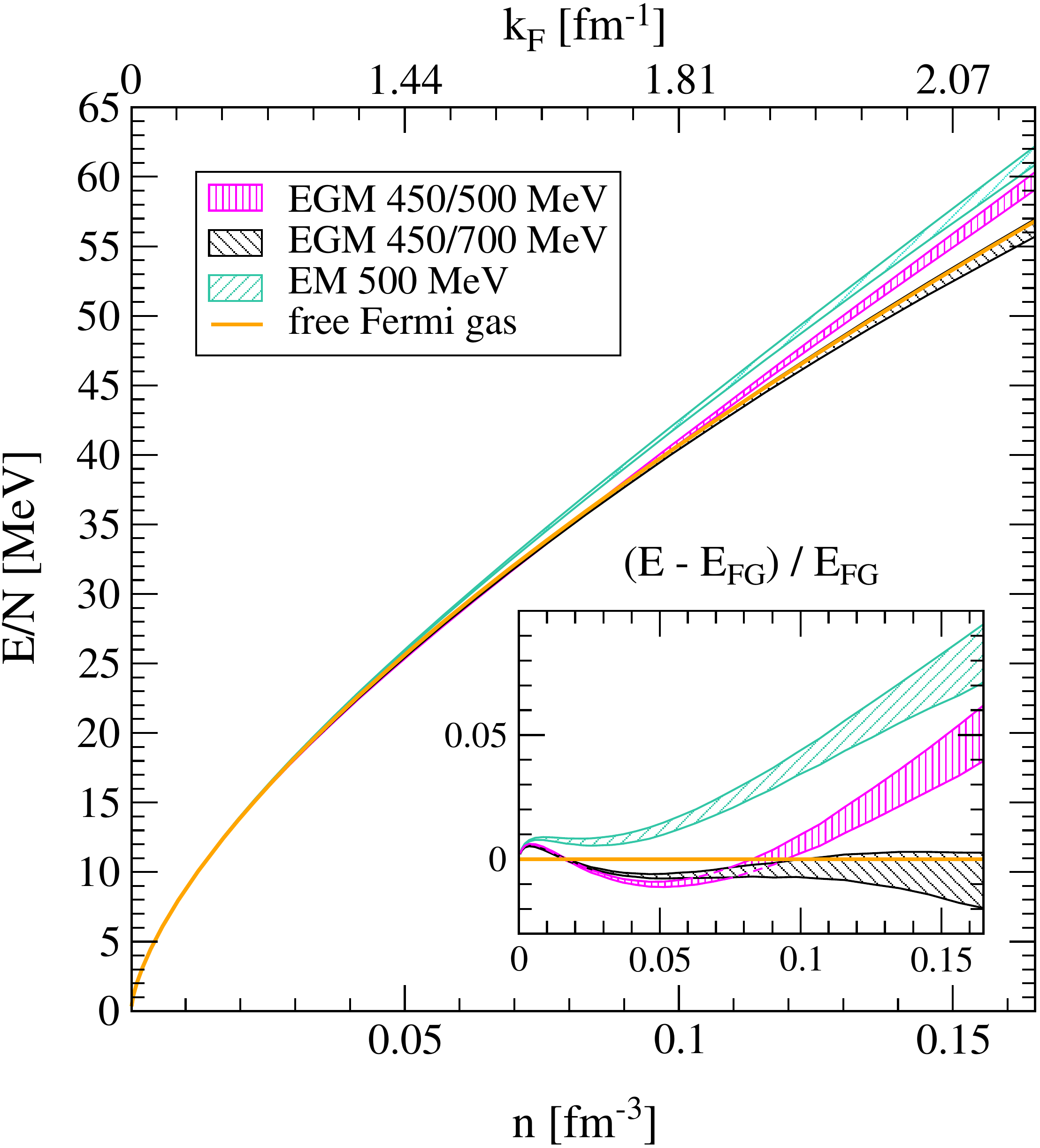}
\end{center}
\caption{(Color online)
Energy per particle of spin-polarized neutron matter
at N$^3$LO as a function of density for the different EM/EGM
NN potentials and including 3N and 4N interactions. The bands provide
an estimate of the uncertainty in 3N forces and in the many-body
calculation (see text). The solid (orange) line is the energy
of a free Fermi gas (FG). The inset shows the relative size of 
the interaction contributions.\label{fig:sumN3LO}}
\end{figure}

In this Letter, we study the properties of spin-polarized neutron
matter at N$^3$LO in chiral EFT~\cite{Epel09RMP}, including
consistently two- (NN), three- (3N), and four-neutron (4N)
interactions. Spin-polarized neutron matter may exist in very strong
magnetic fields as they occur in the interior of magnetars. For a
unitary Fermi gas, the spin-polarized system is a non-interacting gas,
so we ask the question to which densities spin-polarized neutrons
behave like a weakly interacting Fermi gas?  While the answer is
simple at low densities relevant to ultracold atoms, because $P$-wave
interactions between neutrons are weaker and many-body forces are
suppressed by a power of the density, we find the surprising result
(see Fig.~\ref{fig:sumN3LO} for a preview) that the energy of
spin-polarized neutrons is close to a non-interacting system at least
up to saturation density $n_0 = 0.16 \fmiq$, which is well beyond the
large S-wave scattering-length regime $n \lesssim n_0/100$.
Spin-polarized neutron matter has been studied before, e.g., in
Refs.~\cite{Vida02spinpol,Bomb06spinpol,Samm06polDBHF,Bord08polasym},
however with NN interactions only, and without a focus on the subnuclear
density region and the comparison to a weakly interacting Fermi gas.

The physics of spin-polarized neutron matter is interesting, because
it can provide an additional anchor point for energy-density
functionals. To this end, we explore how our results compare with
state-of-the-art functionals. In addition, spin-polarized matter is
ferromagnetic, so that its energy compared to the spin-symmetric
system determines whether a ferromagnetic transition in neutron stars
is possible~\cite{Vida02spinpol,Bomb06spinpol}. Finally, there are
fewer non-trivial contractions for spin-polarized neutrons, so that
the determination of this system is easier in lattice QCD than
symmetric matter~\cite{Bean10NPLQCD}. Therefore, we also study how our
results depend on the pion mass and provide predictions that can be
tested and refined with lattice QCD.

\begin{figure*}[t]
\begin{center}
\includegraphics[width=1.9\columnwidth,clip=]{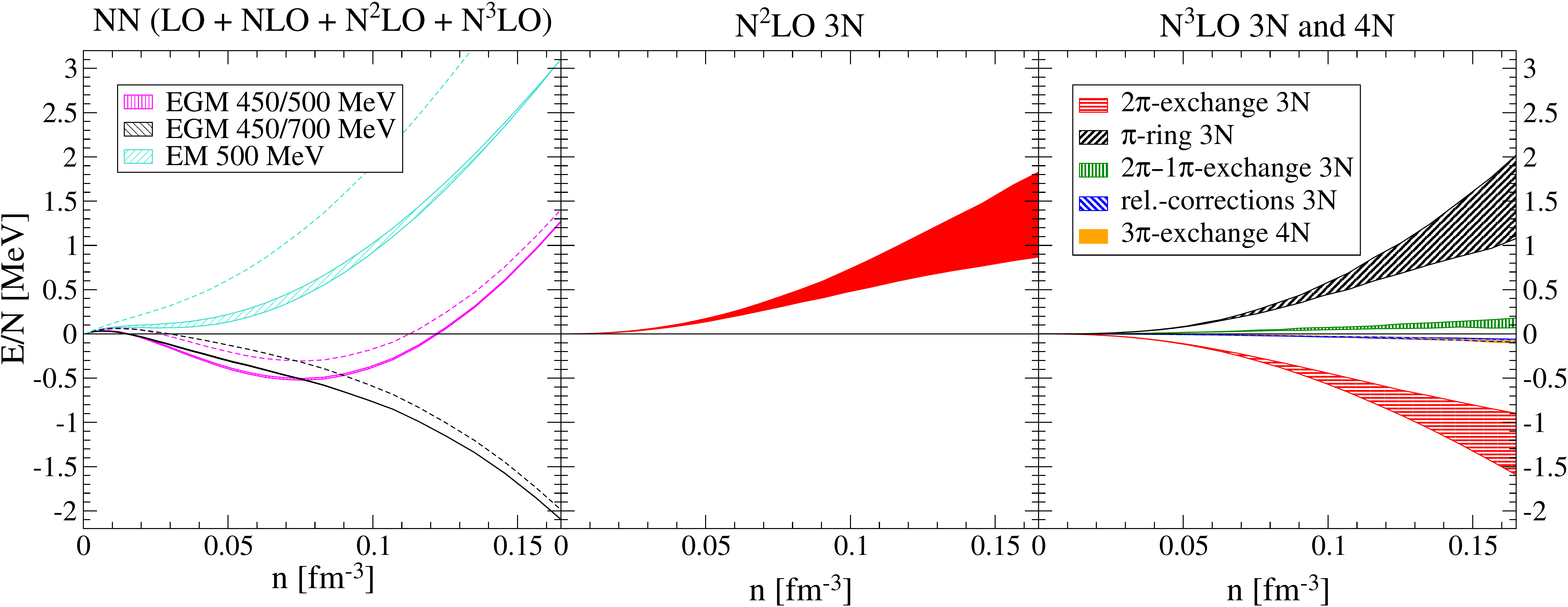}
\end{center}
\caption{(Color online)
Interaction contributions at N$^3$LO to the energy per
particle of spin-polarized neutron matter as a function of
density. The left panel shows the NN contributions for the three NN
potentials. The width of the bands is given by the difference
between second- and third-order contributions in the many-body
calculation. The dashed lines are the Hartree-Fock energies. The
middle panel shows the contribution from N$^2$LO 3N forces, where the
band corresponds to the range of $c_i$ couplings used
and the 3N cutoff variation $\Lambda = 2-2.5\fmi$. The right panel
gives the different N$^3$LO 3N and 4N contributions, with corresponding
$c_i$ and cutoff variations. The 4N contributions overlap with the
relativistic-corrections 3N energies.\label{fig:individual}}
\end{figure*}

\textit{Calculational details.}-- We employ the N$^3$LO NN potential
of Entem and Machleidt (EM) with a cutoff
$500\mev$~\cite{Ente03EMN3LO}, and the potentials developed by
Epelbaum, Gl\"ockle, and Mei{\ss}ner (EGM) with cutoffs $\Lambda /
\widetilde{\Lambda} = 450/500$ and
$450/700\mev$~\cite{Epel05EGMN3LO}. In this way, we
explore a natural cutoff range in Weinberg's power-counting scheme,
although these potentials are not renormalizable for higher
cutoffs~\cite{Nogg05renorm}. In
Refs.~\cite{Tews13N3LO,Krue13N3LOlong}, it was found that
the employed potentials are perturbative in neutron
matter as a result of weaker tensor forces among neutrons and
restricted phase space due to Pauli blocking at finite
densities. This has also been nonperturbatively
verified with quantum Monte Carlo calculations using local chiral
potentials~\cite{Geze13QMCchi}.  Spin-polarized matter is expected
to converge even faster, because $S$-wave interactions among polarized
neutrons vanish, $P$-wave interactions are weaker, and Pauli blocking
becomes even more effective due to the larger Fermi momentum for a
given density compared to spin-symmetric matter.

We include all NN contributions up to second order in many-body
perturbation theory, as well as particle-particle/hole-hole diagrams
to third order (see Ref.~\cite{Hebe10nmatt}). Restricting all spins to
the same spin state, the second-oder contribution to the energy per
particle is given by
\begin{align}
\frac{E_\text{NN}^{(2)}}{N} &=  \frac{1}{4} \left[\prod_{i=1}^4
\int \frac{\mathrm{d}\mathbf{k}_i}{(2\pi)^3}\right]
\frac{|\bra{12} V_{\text{NN}} \ket{34}|^2(2\pi)^3}
{\varepsilon_{\mathbf{k}_1}+\varepsilon_{\mathbf{k}_2}
-\varepsilon_{\mathbf{k}_3}-\varepsilon_{\mathbf{k}_4}}
n_{\mathbf{k}_1}n_{\mathbf{k}_2}\nonumber\\
& \times 
(1-n_{\mathbf{k}_3})(1-n_{\mathbf{k}_4})
\delta(\mathbf{k}_1+\mathbf{k}_2-\mathbf{k}_3-\mathbf{k}_4)\,,
\end{align}
where $n_\mathbf{k}$ denotes the Fermi distribution function at zero
temperature and we use the short-hand notation $i \equiv \mathbf{k}_i$
in the bra and ket states. Taking a free or a Hartree-Fock
spectrum for the single-particle energies $\varepsilon_{\bf k}$
changes the results only at the $10\kev$ level. This indicates that
the many-body calculation is very well converged, and in the following
results are given with a free spectrum. In order to simplify the
numerical calculations, we average over the angles of initial and
final relative momenta $\mathbf{k}$ and $\mathbf{k}'$:
\begin{align}
\int& \frac{\mathrm{d}\widehat{\mathbf{k}}\mathrm{d}\widehat{\mathbf{k}}'}{(4\pi)^2} \, \bigl| \bra{\mathbf{k}\,S=1\,M_S=1}
V_{\text{NN}} \ket{\mathbf{k}'\,S=1\,M_S=1} \bigr|^2\nonumber\\
&=\sum_{l,l',J,\widetilde{J}} 4 (4\pi)^2 C_{ll'}^{J\widetilde{J}}
\bra{k}V_{1l'lJ}\ket{k'} \bra{k'}V_{1ll'\widetilde{J}}\ket{k}\,,
\end{align}
where $V_{Sll'J}$ denote the neutron-neutron partial-wave matrix
elements and $C_{ll'}^{J\widetilde{J}}$ is the sum of
Clebsch-Gordan coefficients $\mathcal{C}_{l_1 m_1 l_2 m_2}^{l_3 m_3}$
\begin{align}
C_{ll'}^{J\widetilde{J}}\! =\! \sum_{M} \mathcal{C}_{l'(M-1)11}^{JM}
\mathcal{C}_{l(M-1)11}^{JM} \mathcal{C}_{l(M-1)11}^{\widetilde{J}M}
\mathcal{C}_{l'(M-1)11}^{\widetilde{J}M}\,.
\end{align}
The angular-averaging approximation has been demonstrated to be
reliable for spin-symmetric matter~\cite{Hebe10nmatt} and
only affects the small contributions beyond Hartree-Fock.

The energy contributions from 3N and 4N forces up to
N$^3$LO~\cite{Kolc94fewbody,Bern083Nlong,Bern113Nshort,Epel064N,Epel074Ndetail} are
calculated in the Hartree-Fock approximation, following the strategy
used in Refs.~\cite{Tews13N3LO,Krue13N3LOlong}.  We expect this
approximation to be reliable since we found only small contributions
from 3N forces at second and third order in perturbation theory in
spin-symmetric neutron matter~\cite{Hebe10nmatt}. In the polarized
case we expect even smaller contributions due to the enhanced Pauli
blocking effects. In addition to the 3N and 4N topologies that do not
contribute to the neutron-matter energy (see
Ref.~\cite{Hebe10nmatt,Krue13N3LOlong}) for the spin-polarized system
also the 3N N$^3$LO two-pion-exchange--contact topology vanishes, as a
consequence of the Pauli principle excluding all leading-oder NN
contacts $C_S$ and $C_T$. Further, the 4N N$^3$LO diagrams $V^e$ and
$V^f$ (according to the nomenclature in Ref.~\cite{Epel064N}) do not
contribute in polarized matter. The $C_S$/$C_T$ dependence of the 3N
N$^3$LO relativistic-corrections interaction is negligible and results
only in energy differences at the $1\kev$ level at saturation
density. Thus, the many-body forces essentially depend only on the
low-energy couplings $c_1$ and $c_3$, which are chosen according to
Ref.~\cite{Rent03ciPWA,Kreb123Nlong}: $c_1 = -(0.75 - 1.13)\gevi$ and $c_3 =
-(4.77 - 5.51)\gevi$ as in Refs.~\cite{Tews13N3LO,Krue13N3LOlong}. In
order to probe the cutoff dependence of our calculation we also vary
the 3N/4N cutoff $\Lambda = 2 - 2.5 \fmi$.

\textit{Results and discussion.}-- Our central result,
Fig.~\ref{fig:sumN3LO}, shows that the energy of spin-polarized
neutrons is close to a non-interacting system, with interaction
effects providing less than 10\% corrections at $n_0$ (see the
inset). The largest dependence of our calculations is on the NN
interaction used. The EM 500 MeV potential leads to weakly repulsive
interactions with $E/N \approx 61.5\mev$ at $n_0$, compared to $55.7
\mev$ for a free Fermi gas. Using the EGM 450/500 and 450/700 MeV
potentials results in even weaker interactions with $E/N \approx 59.5\mev$
and $\approx 56\mev$, respectively. Because $n_0$ for polarized matter
corresponds to a high Fermi momentum of $2.1 \fmi$, these small
differences are due to the range in NN scattering predictions at
these higher momenta.

At very low densities, we can also compare our results to the
dilute-gas expansion~\cite{Hamm00EFT}, where the first non-vanishing
contribution is at $k_{\rm F}^5$ from the P-wave scattering length
$a_{\rm P}$, or the P-wave scattering volume~$a_{\rm P}^3$. We have
fitted the P-wave scattering length for $k_{\rm F} < 0.3$~fm$^{-1}$)
to our equation of state and obtain a range $a_{\rm P}=0.50 - 0.52$~fm
depending on the NN interaction used. This is consistent with
$a_P=0.44 - 0.47$~fm from the different NN interactions with small
corrections due to Pauli blocking that render the P-wave scattering
length more repulsive in the medium.

By comparing our results with the corresponding energy range for
spin-symmetric matter, $E/N \approx 14 - 21\mev$ at
$n_0$~\cite{Tews13N3LO,Krue13N3LOlong}, it is clear that a phase
transition to the ferromagnetic state is not possible for $n \lesssim
n_0$. Further, we expect the energy of spin-polarized neutrons at
higher densities to lie above the free Fermi gas due to repulsive 3N
forces (see also Fig.~\ref{fig:individual}). Assuming the energy of
spin-polarized neutrons remains close to a free Fermi gas also for
higher densities, we can use the general equation of state constraints
of Ref.~\cite{Hebe13ApJ} to provide constraints for the onset of a
possible ferromagnetic phase transition. Taking the three
representative equations of state~\cite{Hebe13ApJ}, a phase transition
to a ferromagnetic state may be possible for $n/n_0 \gtrsim 6.1$, 3.4,
and 2.3 for the soft, intermediate, and stiff equations of state,
respectively. Note that if more massive neutron stars are discovered,
e.g., with $2.4 M_\odot$, the soft case is ruled out~\cite{Hebe13ApJ}.

Figure~\ref{fig:individual} shows the individual interaction
contributions. All energies are small compared to the spin-symmetric
system~\cite{Tews13N3LO,Krue13N3LOlong}. The left panel shows the NN
contributions for the three N$^3$LO potentials. The different behavior
can be traced to different predictions for the scattering phase
shifts. The EM 500 MeV potential gives a net repulsive contribution,
with $E/N \approx 3.1\mev$ at $n_0$ ($5.6 \%$ relative to $E_{\rm
FG}$). Up to densities $n \lesssim 0.1 \fmiq$ the EGM 450/500 and
450/700 MeV potentials are in good agreement and provide only $E/N
\approx - 0.5 \mev$ at $n = 0.08\fmiq$, and then start to differ. The
middle panel of Fig.~\ref{fig:individual} shows the contributions from
the leading N$^2$LO 3N forces. The 3N interactions are, as in the
spin-symmetric case, repulsive but with much smaller energies in the
range $0.8 - 1.9 \mev$ at $n_0$. In the right panel, we show all
contributions from the N$^3$LO many-body forces. The dominant
contributions are from two-pion-exchange 3N forces with energies
$-(0.9 - 1.6)\mev$ at $n_0$. This is almost as large as the leading
contribution of the two-pion-exchange topology, and shows that one is
pushing the chiral EFT expansion to the limits. However, all these 3N
contributions are still small. In addition, there are repulsive
contributions from pion-ring 3N forces, which contribute $1.1 -
2.1\mev$ at $n_0$ and counteract these. Finally, there are small
repulsive contributions from the two-pion--one-pion-exchange 3N
topology of $0.1 - 0.2\mev$ at $n_0$, small attractive contributions
from the relativistic-corrections 3N topology, while three-pion-exchange
4N interactions contribute only $-0.1\mev$ at $n_0$. In total, the
3N+4N contributions provide a net repulsion of $E/N = 1 - 2.2\mev$ at
$n_0$. While it is known that P-wave interactions are 
weak it is remarkable that even contributions from many-body forces are
small.

\begin{figure}[t]
\begin{center}
\includegraphics[width=0.9\columnwidth,clip=]{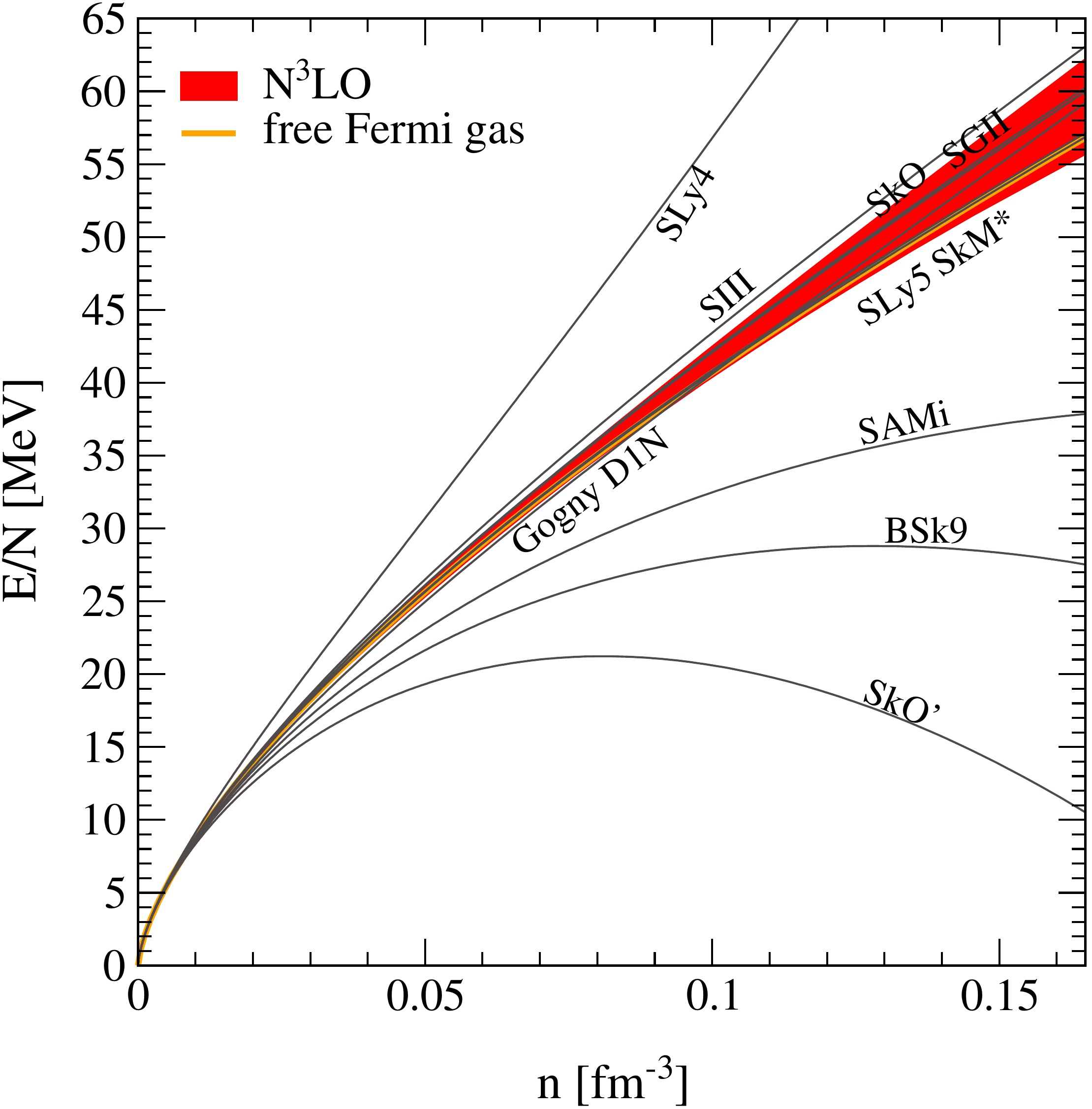}
\end{center}
\caption{(Color online)
Energy per particle of spin-polarized neutron
matter from Fig.~\ref{fig:sumN3LO} in comparison with different
energy-density functionals (see text) following
Ref.~\cite{Forb13polaron}.\label{fig:compEDF}}
\end{figure}

In Fig.~\ref{fig:compEDF} we compare our results with predictions
based on state-of-the-art energy-density functionals 
(for early work on polarized neutron matter with Skyrme functionals
see Ref.~\cite{Kuts93SkyrmeSI}), following
Ref.~\cite{Forb13polaron}: SIII~\cite{Bein75SIII},
SGII~\cite{Giai81SGII}, SkM*~\cite{Bart82SkM}, SLy4 and
SLy5~\cite{Chab98SLy}, SkO and SkO'~\cite{Rein99SkO},
BSk9~\cite{Gori05BSk9}, as well as SAMi~\cite{Roca12SAMi} and using
the Gogny D1N interaction~\cite{Chap08D1N}. At low densities $n
\lesssim 0.01 \fmiq$ all functionals agree with a free Fermi gas.
However, at higher densities we find significant deviations. In best
agreement with our calculations are the functionals SIII, SkO, SGII,
SkM*, and SLy5, whereas the latter two reproduce the free Fermi gas
and the former provide small repulsive contributions. The predictions
of the functionals SLy4, SAMi, BSk9, and SkO' differ significantly
from our N$^3$LO bands. Therefore, it will be interesting to use our
results as additional neutron-matter constraint for modern functionals.
Note that the above discussion of a possible transition to a ferromagnetic
state is different to the spin instabilities caused by the
polarized system to decrease unphysically in energy, as for the SkO' case.

\begin{figure}[t]
\begin{center}
\includegraphics[width=0.9\columnwidth,clip=]{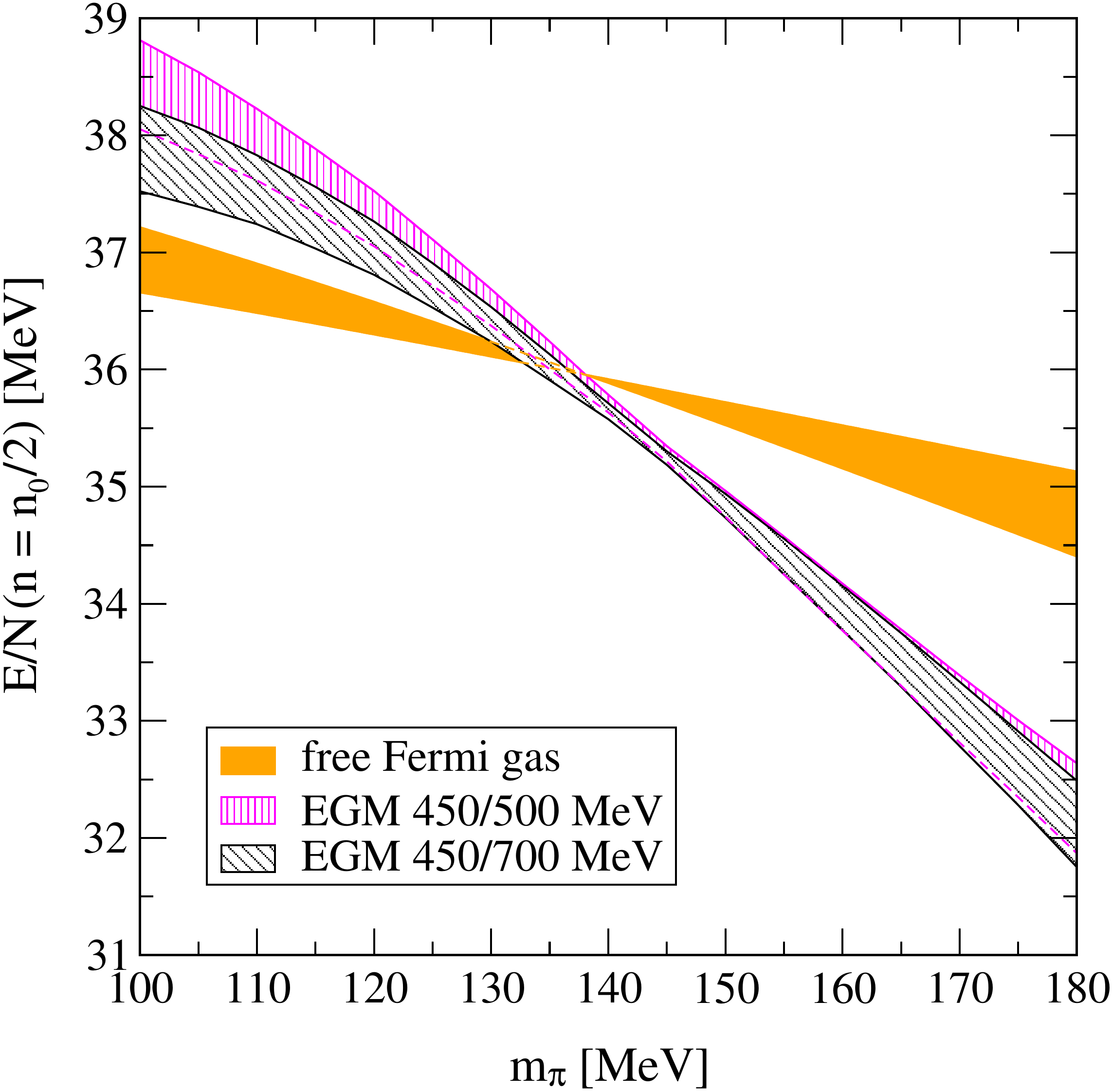}
\end{center}
\caption{(Color online)
Energy per particle of spin-polarized neutron 
matter at a density $n = n_0/2$ as a function of the pion mass. The
solid (orange) band indicates the energy of a free Fermi gas. The
shaded bands correspond to the EGM NN potentials, including 3N and
4N interactions, at the same many-body calculational level as the
results in Fig~\ref{fig:sumN3LO}.\label{fig:mpivar}}
\end{figure}

For comparison with lattice QCD simulations, we also vary the pion
mass in NN, 3N, and 4N interactions. For this estimate we only take
into account the explicit pion exchanges and do not vary the pion mass
implicitly in the coupling constants. For spin-symmetric neutron
matter, this was found to be the dominant contribution,
whereas the contributions from the pion-mass dependence of
the coupling constants was estimated to be smaller~\cite{Kais09condnm,Laco10cond,Krue13chicond}. The
dependence of the energy of the free Fermi gas, $E_\text{FG}/N = 3
k_\mathrm{F}^2/(10m_N)$, is a result of the change of the nucleon mass
with the pion mass. This varies as~\cite{Bere13mqdep}
\begin{align}
m_N (m_{\pi}) = m_0 - 4c_1 m_\pi^2 - \frac{3g_A^2}{32\pi f_\pi^2} m_\pi^3
+ \mathcal{O}(m_\pi^4)\,,
\end{align}
where $m_0$ is the nucleon mass in the chiral limit and $c_1$ is the
same low-energy coupling that enters NN and 3N forces at N$^2$LO. We
consistently also do not include the implicit pion-mass dependence
of the coupling-constants in these estimates. For $c_1$ we use the
same range above, as in the 3N forces. Using the physical values of $m_N$,
$m_\pi$, $g_A$, and $f_\pi$ we can extract $m_0$ for the employed
$c_1$ range. This leads to the filled (orange) band in
Fig.~\ref{fig:mpivar} at $n_0/2$ and corresponds to a range of the
pion-nucleon sigma term $\sigma_{\pi N} = 34.9 - 63.9\mev$. In
Fig.~\ref{fig:mpivar} we show that including interactions gives a very
similar $m_\pi$ dependence, but away from the physical pion mass, the
energy starts to deviate more from the free Fermi gas.
As in spin-symmetric matter, the interaction contributions also
increase the chiral condensate, as determined from the slope in
$m_\pi$. We emphasize that for a precision comparison, one also needs
to include the $m_\pi$ dependence of the low-energy couplings in
nuclear forces.

\textit{Summary and outlook.}-- We have presented a complete N$^3$LO
calculation of spin-polarized neutron matter, where the dominant
uncertainty is due to the NN potential used, as well as due to the
uncertainty in 3N forces. The uncertainty from the many-body
calculation is very small (shown by the bands in the left panel of
Fig.~\ref{fig:individual}). Our results show that the energy of
spin-polarized neutrons is remarkably close to a non-interacting
system. This shows that the physics of neutron matter is similar to a
unitary gas well beyond the scattering-length regime. Moreover, our
results provide constraints for energy-density functionals of nuclei
and show that a phase transition to a ferromagnetic state is not
possible for $n \lesssim n_0$. Finally, our predictions can be tested
and refined with lattice QCD calculations of spin-polarized neutrons
in a box.

\begin{acknowledgments}

We thank W.~Detmold, C.~Drischler, T.~Lesinski, M.~Savage, and I.~Tews
for useful discussions. This work was supported by the DFG through
Grant SFB 634, the ERC Grant No.~307986 STRONGINT, and the Helmholtz
Alliance Program of the Helmholtz Association, contract HA216/EMMI
``Extremes of Density and Temperature: Cosmic Matter in the
Laboratory''. We thank the Institute for Nuclear Theory at the
University of Washington for its hospitality and the DOE for partial
support during the completion of this work.

\end{acknowledgments}

\bibliography{strongint}
\bibliographystyle{apsrev}

\end{document}